\documentclass{PoS}
\usepackage{epsfig}
\usepackage{graphicx}

\title{Electric Polarizabilities from Lattice QCD}

\ShortTitle{Electric Polarizabilities from Lattice QCD}

\author{W.~Detmold\\
    	Department of Physics\\	
	University of Washington\\
	Box 351560\\
	Seattle, WA 98195-1560, USA\\
	E-mail: \email{wdetmold@phys.washington.edu}
}

\author{\speaker{B.~C.~Tiburzi}\\
        Maryland Center for Fundamental Physics\\
        Department of Physics\\
        University of Maryland\\
        College Park, MD 20742-4111, USA\\
        E-mail: \email{bctiburz@umd.edu}
        }

\author{A.~Walker-Loud\\
        Department of Physics\\
        College of William and Mary\\
        Williamsburg, VA 23187-8795, USA\\
        E-mail: \email{walkloud@wm.edu}
        }

\abstract{
The response of hadrons to electromagnetic probes is highly constrained by chiral dynamics; 
but, in some cases, 
predictions have not compared well with experimental data. 
The lattice can be used
to test the chiral electromagnetism of hadrons and ultimately confront experiment.
We use background field techniques to study the electromagnetic polarizabilities of hadrons. 
Focusing on simulations in background electric fields, we present preliminary results for both charged and neutral particle polarizabilities. 
The former are extracted using a novel method.}

\FullConference{The XXVI International Symposium on Lattice Field Theory\\
		 July 14-19 2008\\
		 Williamsburg, Virginia, USA}

\begin{document}

\section{Introduction}

The study of QCD with external sources provides 
a controllable aspect to the non-perturbative dynamics;
and, in turn, allows for us to glean information about 
the quark and gluon structure of hadrons. 
The quarks possess electric charges; and, consequently 
hadrons polarize in applied electric and magnetic fields. 
This is despite the considerably stronger chromodynamic
forces that confine quarks and gluons into hadrons. 
Aside from Born couplings to the hadron's total charge, 
the effective Hamiltonian of a hadron in external
fields is given by: 
$H = - \frac{1}{2} \alpha_E \vec{E}^2 - \frac{1}{2} \beta_M \vec{B}^2$.
This Hamiltonian arises as a low-energy approximation. 
One can consider further terms with more derivatives, 
or terms with higher powers of the field strength. 
Working with constant fields of sufficiently small size, 
we can neglect such higher-order terms.%
\footnote{
Additionally this effective Hamiltonian is valid in the infinite
volume limit. Further terms must be added to account for 
infrared effects that stem from boundary conditions on a compact space~\cite{Hu:2007eb}.
}
The coefficients of the operators in the effective Hamiltonian are the electric polarizability 
$\alpha_E$, 
and magnetic polarizability 
$\beta_M$. 
The long-range structure of hadrons is dominated by pion interactions. 
If we imagine a typical hadron as a core surrounded by a virtual pion cloud, 
then the charged pion cloud polarizes in external electromagnetic fields. 
More precisely, the form of the electric polarizability, for example, is highly 
constrained by chiral dynamics. For any hadron $h$, the leading-order
electric polarizability has the form:
$\alpha^h_E = N^h  \frac{e^2}{m_\pi \Lambda_\chi^2}$,
where $e$ is the electric charge, $\Lambda_\chi$ is the chiral symmetry 
breaking scale, and $N^h$ is a pure number and depends upon the 
particular hadron. 
Curiously, polarizabilities are singular in the $SU(2)$ chiral limit as they are proportional 
to the inverse pion mass.%
\footnote{
The only exceptions are the kaons and eta. Their polarizabilities scale
as $m_K^{-1}$ and $m_\eta^{-1}$, respectively, and are only singular 
in the $SU(3)$ chiral limit. 
}

Experimentally polarizabilities can be measured from low-energy 
Compton scattering experiments off hadronic targets, for a review see~\cite{HydeWright:2004gh}. 
This is, of course, only directly possible for the proton.
There are a number of areas where our current understanding
is limited:
experimental determinations of pion polarizabilities have always been at odds
with chiral perturbation theory, 
neutron polarizability measurements are indirect using
Compton scattering off light nuclei (largely deuterium)
which must consequently rely on theoretical input,   
extraction of the nucleon magnetic polarizability from 
experiment currently has $\sim 50 \%$ error bars, 
and finally two combinations of spin-polarizabilities have yet to be measured. 
The situation is likely to change in the near future:
new pion and first kaon polarizability results are expected
from COMPASS at CERN, 
Compton scattering experiments off deuterium
are being performed at MAX-lab in Lund, 
and 
high precision results for all nucleon polarizabilities 
are anticipated in the next few years from the high intensity gamma source at TUNL. 
It is our hope that the lattice can play a timely phenomenological role
with respect to hadronic polarizabilities.

\section{Lattice Details}

While polarizabilities enter in lattice four-point functions, 
that method of extraction is not feasible, for example, 
due to the need for rather long momentum extrapolation.  
We have chosen to work with classical external electromagnetic fields.  
Currently we have artificially set the sea quark charges to zero, and 
thus calculate the effects of electromagnetism coupled to valence 
quarks only. 
There are a number of ways in which physical predictions can be 
made from this scenario; but, they all hinge of the applicability of  
the low-energy effective theory.
We use RBC/UKQCD domain-wall fermion gauge configurations~\cite{Allton:2007hx}, 
the ensembles are summarized in Table~\ref{t:ensembles}. 
Computational restrictions require us to choose numerically 
cheap valence quarks due to the requisite number of 
propagator inversions. 
%
%
%
  \begin{table}
    \begin{tabular}{c|ccccc||ccc|c}
    Set & 
    $N_{config}$ & 
    $L^3 \times \beta$ & 
    $m_s$ & 
    $m_u$ &
    $m_\pi$ & 
    $\kappa_s$ & 
    $\kappa_u$ & 
    $\kappa_{critical}$ &
    $!$ \\
    \hline
    \hline
    $1$ &
    $141$ &
    $16^3 \times 32$ &
    $0.04$ &
    $0.01$ &
    $400 \, \texttt{MeV}$ &
    $0.13810$ &
    $0.13939$ &
    $0.14000$ &
    $\sim 10$
\\
    $2$ &
    $228$ &
    $24^3 \times 64$ &
    $0.04$ &
    $0.01$ &
    $420 \, \texttt{MeV}$ &
    $0.13806$ &
    $0.13934$ &
    $0.13993$ &
    $\sim 2$
\\
    $3$ &
    $174$ &
    $24^3 \times 64$ &
    $0.04$ &
    $0.005$ &
    $330 \, \texttt{MeV}$ &
    $0.13811$ &
    $0.13957$ &
    $0.13993$ &
    $\sim 15$
\\
\hline
      \end{tabular}
    \caption{
    Summary of RBC/UKQCD~\cite{Allton:2007hx} domain wall ensembles used, 
    as well as the $\kappa$-parameters for the valence clover fermions, and the number ($!$)
    of exceptional gauge configurations encountered. 
    }
  \label{t:ensembles}
 \end{table}
%
%
We have opted for tadpole-improved clover fermions, 
with plans in the future to repeat our calculation
using domain-wall valence quarks. 
To reduce unitarity violations,
we have tuned the valence and sea pion masses 
to within statistical error (additionally for the $\eta_s$ meson). 
Details of the $\kappa$-tuning are also shown in the table. 
At these comparatively light values of the pion mass, we 
encounter exceptional gauge configurations. This is particularly
troublesome on the smaller volume, and at the lightest pion mass, 
as expected. As a temporary fix, we shall simply drop the exceptionals
and focus on ensemble Set $2$, for which there were only $2$ such configurations. 
If the future, we will HYP-smear to avoid this problem.

\section{Background Fields}

We will focus our attention on background electric field calculations that we have performed. 
To couple a classical electric field to valence quarks, we post-multiply the existing 
color links 
$U_\mu(x)$,  
by the Abelian link 
$U_\mu^{cl}(x)$, 
so that the net effect is 
$U_\mu(x) \longrightarrow U_\mu(x) U_\mu^{cl}(x)$. 
This choice is natural for a gauge invariant lattice theory, but 
not mandated: any sensible way of including background
fields will agree in the continuum limit and hence differ by 
irrelevant operators. 
As we work in Euclidean space, the form we choose 
for the Abelian links is $U_\mu^{cl}(x) = \delta_{\mu 3} \exp( - i q \mathcal{E} x_4 )$. 
To arrive at Minkowski space results, one technically needs
to perform the analytic continuation $\mathcal{E} \to i E$. 
Non-perturbative effects due to the characteristic instabilities 
of an electric field will hence be absent in our formulation. 
This is a feature not a bug: such real time processes
would obstruct our Euclidean space calculation. 
Instead we analytically continue the perturbative
expansion of observables in powers of $\mathcal{E}$. 
The electric polarizability is the first non-vanishing term in this expansion, 
and the perturbative analytic continuation is trivial, see~\cite{Tiburzi:2008ma}.

The implementation of a constant gauge field on a torus requires quantization conditions~\cite{'tHooft:1979uj}. 
One seeks to define a periodic lattice action on a torus, which is a closed surface. 
Thus given the field we wish to put down, consider a region in the $x_3$-$x_4$ 
plane with area $A_1$. 
The flux out of this region is $\exp ( i q \mathcal{E} A_1)$.
The total area of the $x_3$-$x_4$ plane is $\beta L = A_1 + A_2$. 
Because the torus is a closed surface, the flux out of $A_1$ must be 
the flux into $A_2$. 
Hence we find the quantization condition
$ q_d \mathcal{E} = \frac{2 \pi n}{\beta L}$, 
where $q_d$ is the down quark electric charge
and $n$ is an integer. 
Quantizing the flux for down guarantees that the flux for up is quantized.

\begin{figure}
\includegraphics[width=7cm]{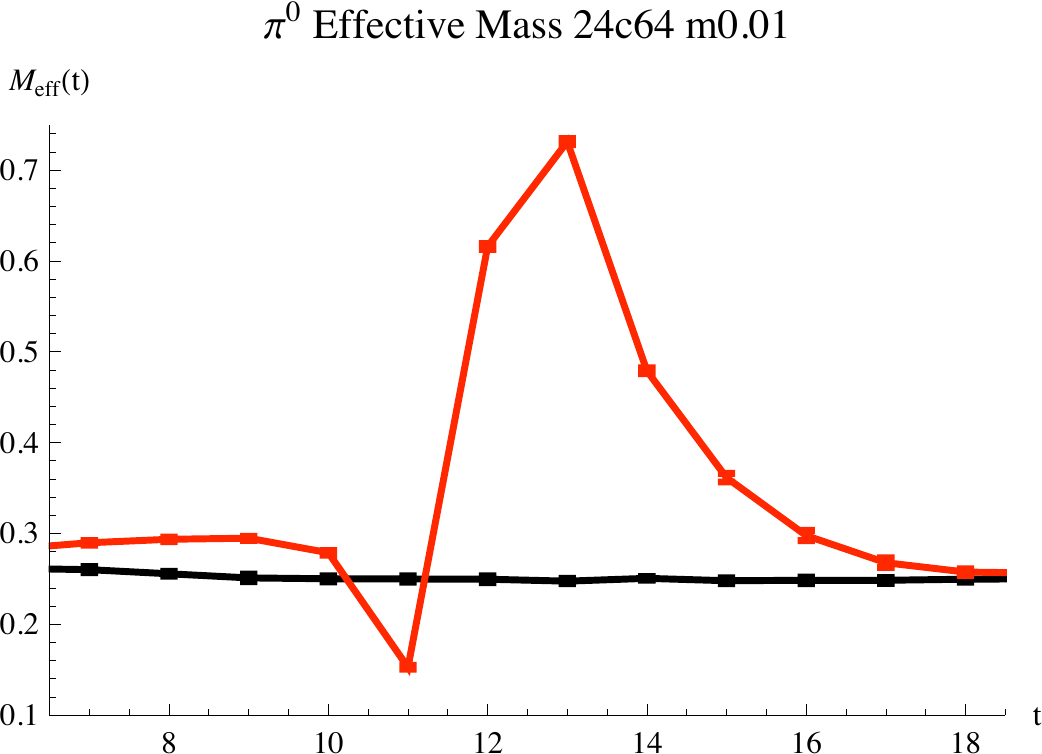}
$\qquad$
\includegraphics[width=7cm]{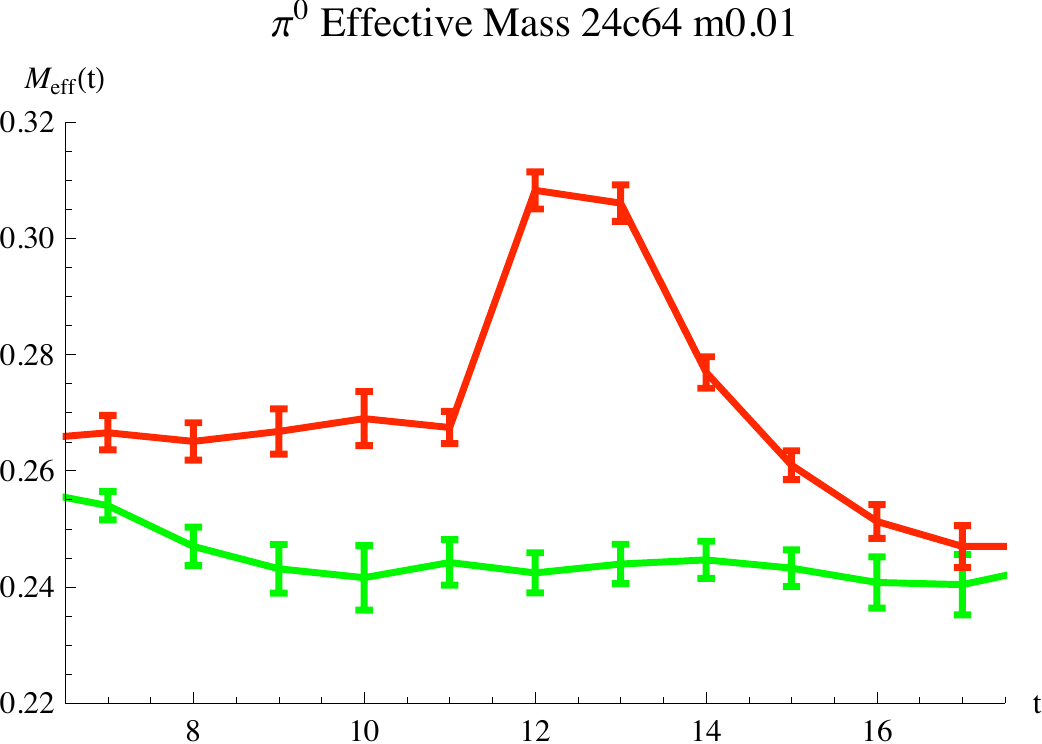}
\caption{\label{f:compare} 
Comparison of background field implementation.
On the left, we use the na\"ive implementation $U_\mu^{cl}(x)$
and show neutral pion effective mass plots for $n=3$.
The bottom curve has $t_{source} = 0$, while the top curve has
$t_{source} = 52$ (so that, after translation, the field spikes at $t=12$ on the plot).
On the right, we use $U_\mu^{cl}(x)$ but include the additional transverse links. 
Shown are effective mass plots with source time $t_{source}=52$ for 
the field strengths $n=3$ and $n=e$ corresponding to the bottom and top curves, respectively.    
In each case, it is the connected part of the neutral pion correlator that was calculated.}
\end{figure}

The above argument is valid for a continuous torus. 
On a discrete torus, the argument must be augmented~\cite{Smit:1986fn}, 
and is only possible using the link formulation to include the gauge field. 
The goal is to render each of the elementary plaquettes in the $x_3$-$x_4$
plane identical, with value: $\exp( i q \mathcal{E})$. 
In the bulk of the lattice, this is already accomplished with $U_\mu^{cl}(x)$. 
There are $L$ plaquettes, however, that are different. 
They each wrap around from $x_4 = \beta -1$ to $x_4 = 0$,
and have the value $\exp [ i q \mathcal{E} (1 - \beta) ]$. 
One can add transverse links at the time boundary to remove the spike. 
This relocates the defect to just one plaquette:
the plaquette starting at the far corner of the $x_3$-$x_4$ plane.
The value of this plaquette is: $\exp [ i q \mathcal{E} ( 1 - \beta L)]$, 
which is uniform precisely when the quantization condition is met.
Modifying the links is equivalent to a gauge transformation, 
but one that is singular in the continuum limit.
We have run propagators using 
various values of quantized and non-quantized field strengths
with differing locations for the source time. 
Shown in Figure~\ref{f:compare} are four different cases. 
%
%
%
%
%
%
%
%
Without transverse links, the figure shows undesirable behavior
due to field strength spike at the boundary. With the addition
of transverse links, only non-quantized field strengths see a 
spike at the boundary. The quantized case corresponds to a 
completely  periodic lattice action.

\section{Selected Results}

\begin{figure}
\includegraphics[width=4.5cm,angle=270]{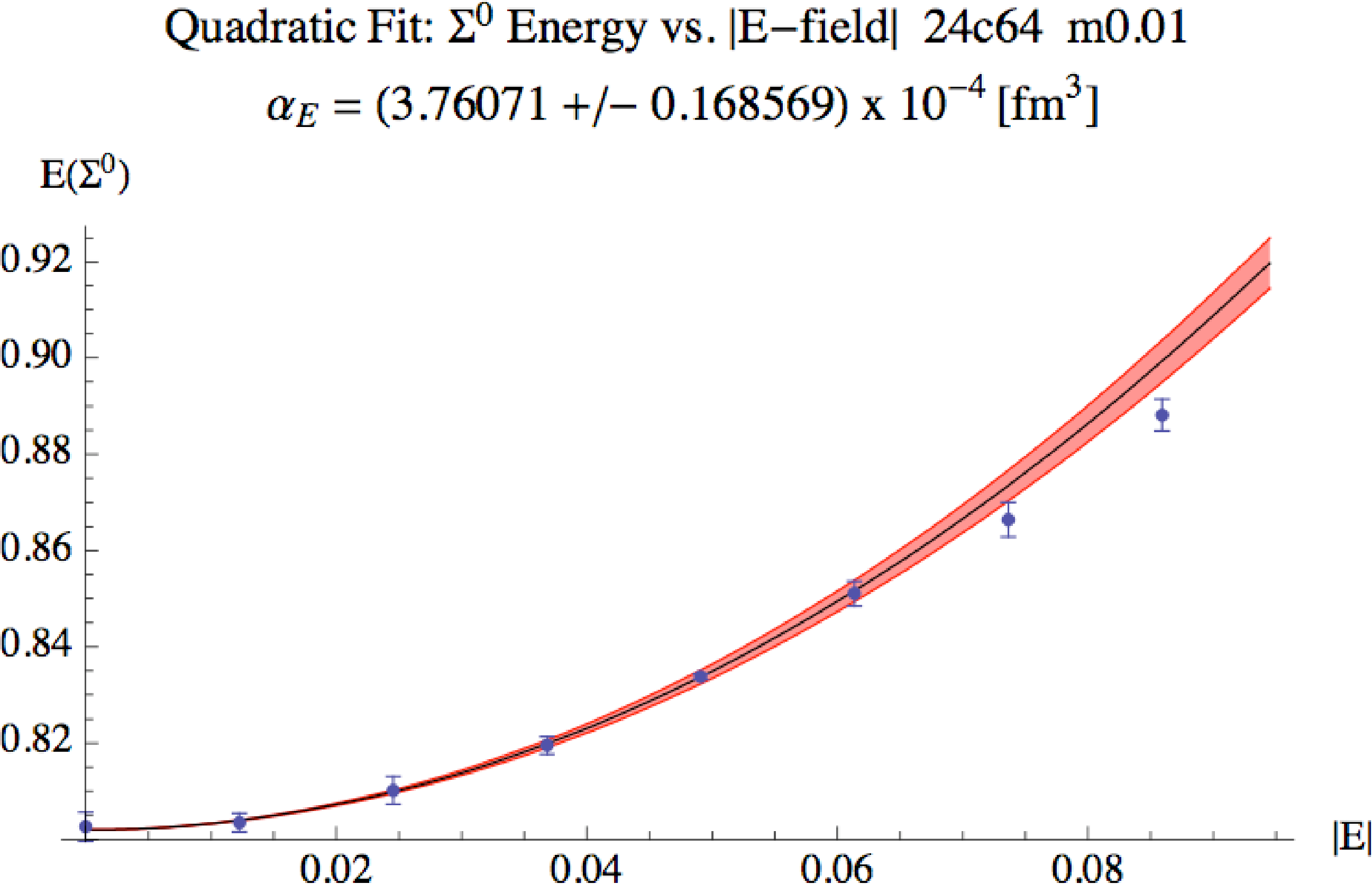}
$\qquad$
\includegraphics[width=4.5cm,angle=270]{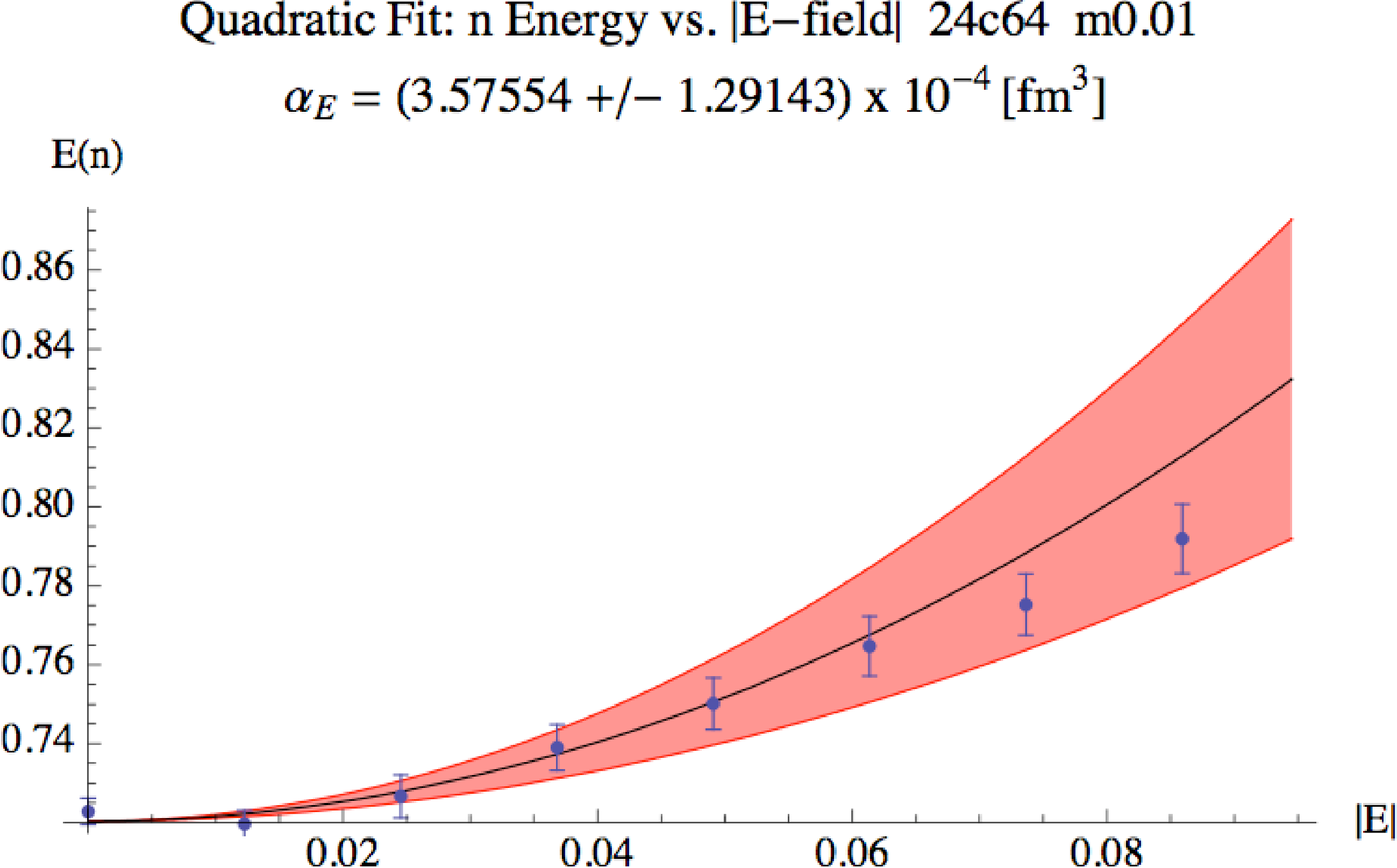}
\caption{\label{f:neutral} 
Plots of extracted energies $E$ versus electric field $|\mathcal{E}|$ for the $\Sigma^0$ and $n$. 
The band corresponds to a one-sigma variation of the extracted value for the polarizability.  
In both cases, the largest two field values have been excluded from the fit. 
Consistent values for the polarizability are extracted when these data are retained
but the energy is fit to a quadratic plus quartic form. 
}
\end{figure}

The measurement of neutral particle electric polarizaiblities can be
done using standard spectroscopy. One measures the long-time
behavior of the two-point correlation function to deduce 
the single-particle energy $E$ in the background field.%
\footnote{
Infrared effects modify this procedure. One must additionally
allow for renormalization of the single-particle effective action.
We have neglected such volume corrections.
} 
This energy has an expansion in powers of the electric field,
$E = M + \frac{1}{2} \alpha_E \mathcal{E}^2 + \frac{1}{4!} \overline{\alpha}_E \mathcal{E}^4 + \ldots$.  
By calculating the energy for a variety of values of the quantized electric field strengths,
we can fit the coefficients $\alpha_E$, and $\overline{\alpha}_E$.  
We use values for $U_\mu^{cl}(x)$ corresponding to 
$n = 1$, $-2$, $3$, $4$, $5$, $-6$, and $7$. 
In Figure~\ref{f:neutral}, we show extracted energies versus electric field
strength for the $\Sigma^0$ and $n$. 
%
%
%
%
%
Not surprisingly the neutron results are noisier. 
Preliminary extracted values for select polarizabilities are as follows
(each given in units of $10^{-4} \, \texttt{fm}^3$):
$\alpha_E^{\pi^0} = 0.12 \pm 0.78$, consistent with expectations
from one-loop chiral perturbation theory for the connected part of the correlation function;
$\alpha_E^{K^0} = 0.27 \pm 0.05$, size consistent with one-loop chiral expectations; 
$\alpha_E^{K^{* 0}} = -1.2 \pm 0.4$, which is surprisingly negative; 
$\alpha_E^n = 3.6 \pm 1.3$; 
$\alpha_E^\Lambda = 3.1 \pm 0.8$; 
$\alpha_E^{\Sigma^0} = 3.8 \pm 0.2$; 
and
$\alpha_E^{\Xi^0} = 2.8 \pm 0.3$.

\begin{figure}
\includegraphics[width=7.1cm]{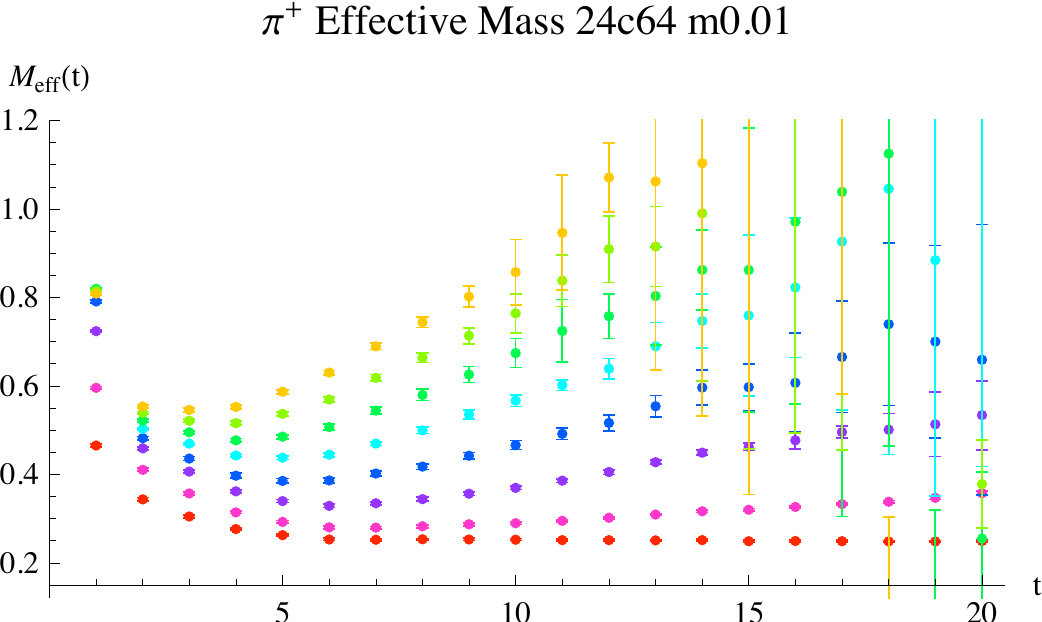}
$\qquad$
\includegraphics[width=7.1cm]{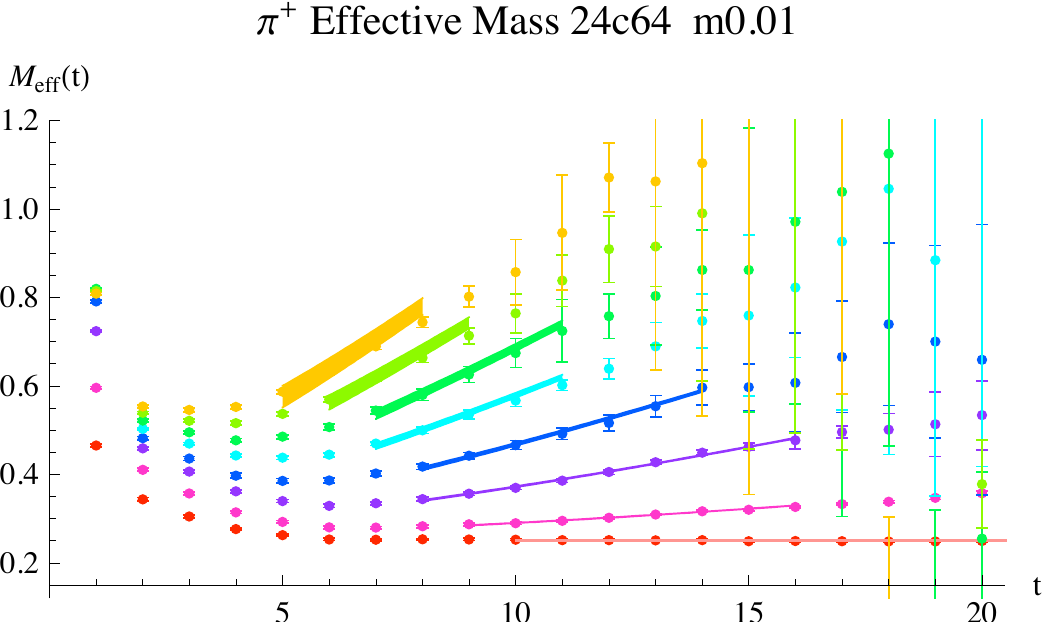}
\caption{ \label{f:PiPlus} 
Plot of the effective mass for the charged pion in electric fields. 
On the left, the effective mass is shown to be both field strength dependent
and non-constant in time. 
On the right, we fit the time dependence using the full functional form of 
$g(E,\mathcal{E},\tau)$.
}
\end{figure}

Charged particle polarizabilities can also be extracted from lattice QCD correlation functions. 
Standard spectroscopy is no avail here as energy is not a good quantum number. 
Instead, one considers the single-particle effective action. 
There are Born and non-Born terms. The non-Born terms can be summed
as in the case of a neutral particle into $E$ defined above. This ordinarily 
would be the energy of the hadron in the external field. 
Additionally one must sum the Born couplings to the particle's total charge.
This leads to modified behavior of the charged particle's two-point function.  
Assuming the ground state hadron dominates over a window of time, we have
\begin{equation} \label{eq:cylinder}
G(\tau)
=
\sum_{\vec{x}} 
\langle 0 |  \chi(\vec{x}, \tau) \chi^\dagger(\vec{0}, 0) | 0 \rangle
=
Z \, g( E, \mathcal{E}, \tau)
.\end{equation}
The function $g$ can be written in terms of parabolic cylinder functions. 
The point being: it is no longer a simple exponential falloff. 
We can gain intuition about $g$ by considering the non-relativistic
limit, in which we have~\cite{Detmold:2006vu}
\begin{equation} \label{eq:NR}
g( E, \mathcal{E}, \tau) 
\longrightarrow
\exp \left(
- E \tau - \frac{Q^2 \mathcal{E}^2 \tau^3}{6 M}
\right)
.\end{equation}
The non-relativistic limit will always be valid for small times 
(but still large enough to filter out the ground state).
On a standard effective mass plot, 
the logarithm of the ratio of correlators will no 
longer plateau, but will rise. 
In the short time limit, Eq.~(\ref{eq:NR}) 
shows $M_{eff}(\tau) = C + \tau(\tau+1) \frac{Q^2 \mathcal{E}^2}{2 M}$. 
We show the behavior of the charged pion effective
mass as a function of time in Figure~\ref{f:PiPlus}. 
%
%
%
%
%
%
%
%
%

\begin{figure}
\includegraphics[width=2.35cm,angle=270]{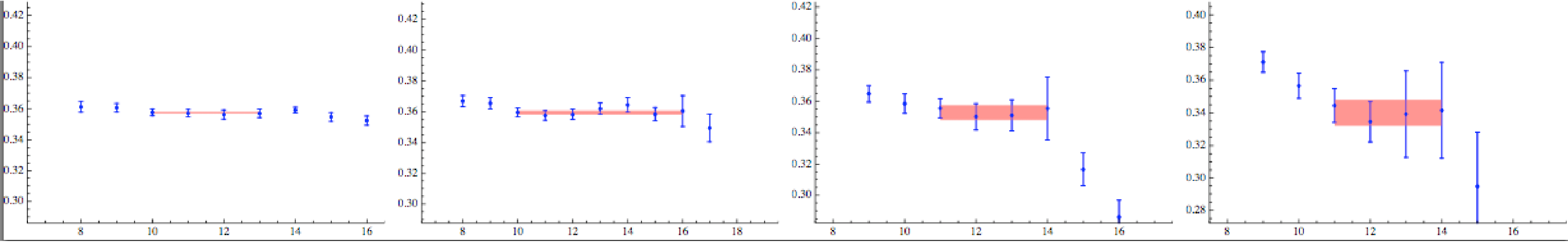}
\caption{ \label{f:EffectiveEnergy} 
Effective energy plots for the $K^+$. 
Plotted are the values of $E_{eff}$ versus time for the 
$n=1$, $-2$, $3$, and $4$ values of the field strength (shown left to right). 
}
\end{figure}

Typically an effective mass plot is used to guide the eye in fitting the correlation function.
For charged particles, the effective mass is no longer an effective tool for this purpose. 
A more reliable plot, we deem the effective energy plot.
To generate values for this plot, we solve for $E_{eff}$ using two time steps
\begin{equation}
\frac{g(E_{eff}, \mathcal{E}, \tau + 1)}{g(E_{eff}, \mathcal{E}, \tau)}
= 
\frac{G(\tau+1)}{G(\tau)}
,\end{equation}
for all values of $\tau$. 
This must be done numerically, and is particularly taxing to generate error bars on $E_{eff}$ at each time slice. 
Fortunately there is a one-dimensional integral representation for $g(E, \mathcal{E}, \tau)$. 
We then search for a plateau for $E_{eff}$ in time. 
Plotted in Figure~\ref{f:EffectiveEnergy} are effective energy plots
for the charged kaon. 
%
%
%
%
%
Guided by the effective energy plot, we fit for $E$ using Eq.~(\ref{eq:cylinder})
for each value of the field strength. We can then plot the would-be energies
as a function of $\mathcal{E}$ and perform fits to extract the coefficient
of the quadratic term. This is how one deduces the electric polarizability
of a charged particle.

\begin{figure}
\includegraphics[width=4.35cm,angle=270]{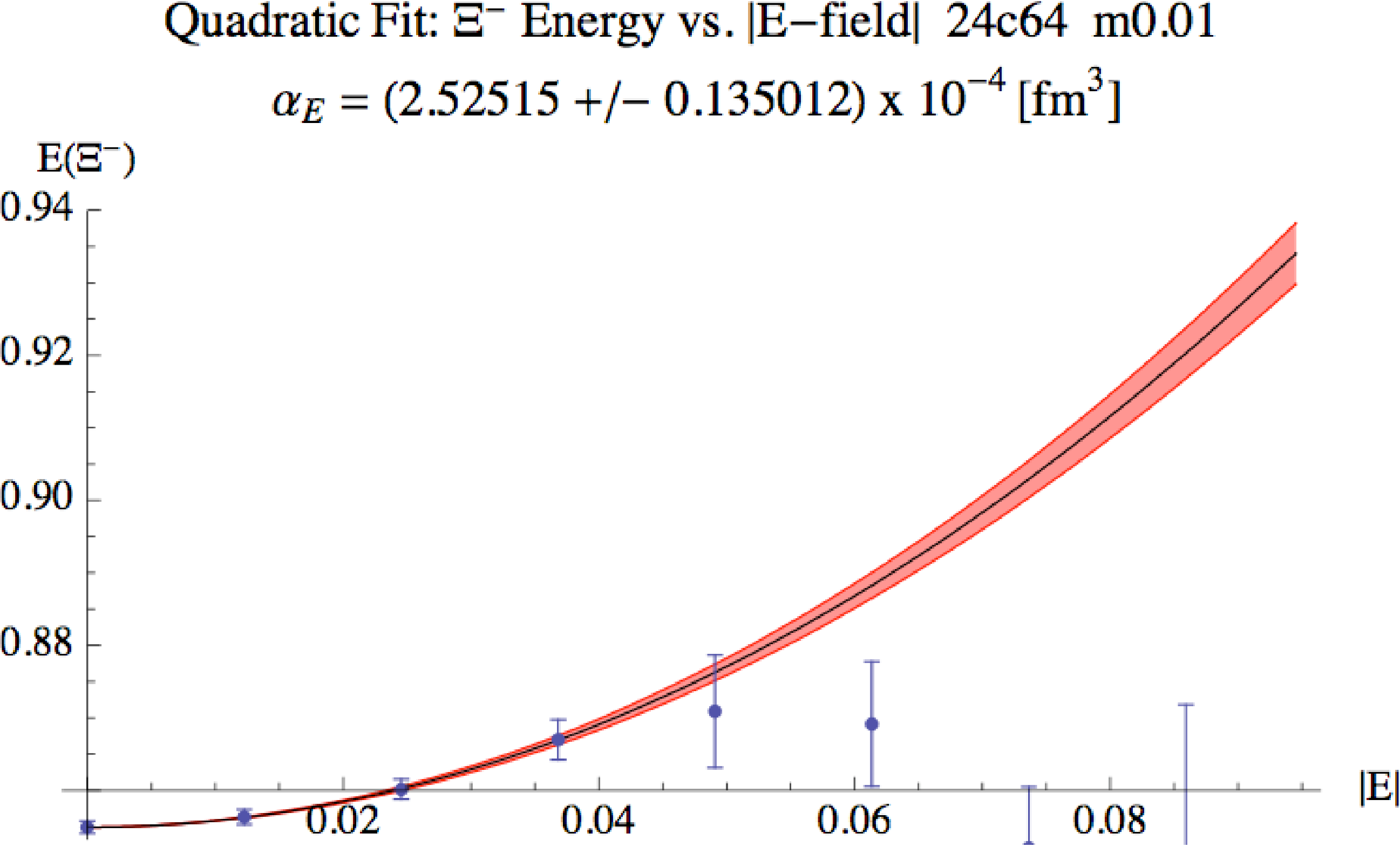}
$\qquad$
\includegraphics[width=4.35cm,angle=270]{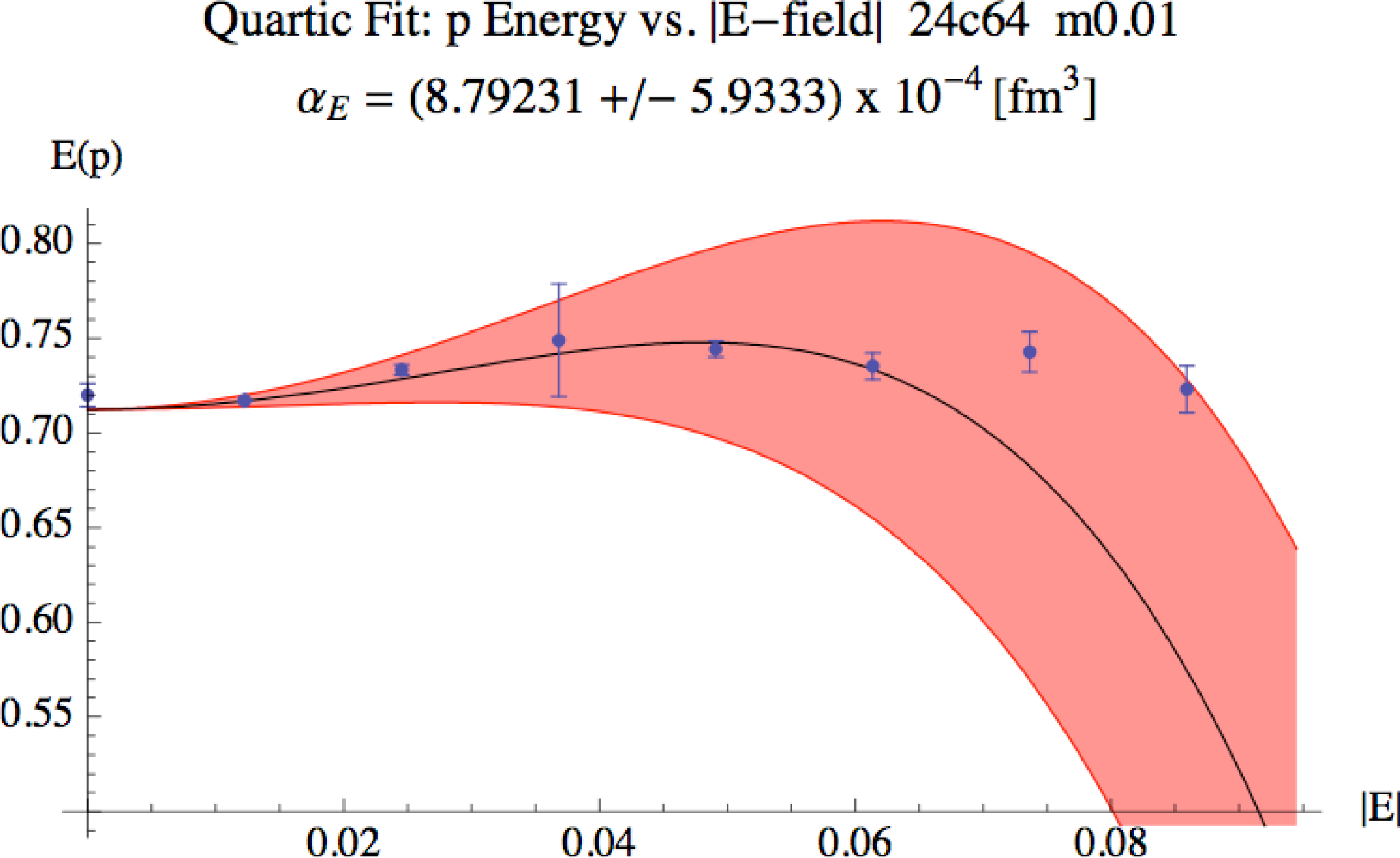}
\caption{\label{f:charged} 
Plots as in Figure~2, but for charged particles: the $\Xi^-$ and $p$. 
In the case of the $\Xi^-$, the highest four field strengths have been 
excluded from the quadratic fit. For the proton, shown is a quadratic
plus quartic fit excluding the two largest field strengths. 
}
\end{figure}

Plotted in Figure~\ref{f:charged} are extracted values of $E$ versus $|\mathcal{E}|$
for the $\Xi^-$ and $p$. The proton is not surprisingly noisier. 
For charged particles, we find a rather ubiquitous feature: 
the coefficient of the quartic $\mathcal{E}^4$ term in the expansion 
of $E$ appears to be larger than that for neutral particles. 
As the strength of the field grows, this quartic term rapidly 
pulls the curve downward making it comparatively more difficult
to extract the quadratic term. We are uncertain about the origin 
of this systematic effect.
Select values for electric polarizabilities of charged particles
are:
$\alpha_E^{\pi^+} = 3.4 \pm 0.4$, 
$\alpha_E^{K^+} = 2.8 \pm 2.2$, 
$\alpha_E^{K^{*+}} = - 5.9 \pm 2.6$, 
$\alpha_E^{p} = 8.8 \pm 5.9 $,
$\alpha_E^{\Sigma^+} = 5.3 \pm 2.2$, 
and
$\alpha_E^{\Xi^-} = 2.5 \pm 0.1$. 
These values are again preliminary and are given in units of $10^{-4} \, \texttt{fm}^3$. 
Despite the large error bar on the $K^+$ polarizability, 
the ratio $\alpha_E^{\pi^+} / \alpha_E^{K^+}$ is in good agreement
with expectations from one-loop chiral perturbation theory, 
namely this ratio approximately scales as $m_K / m_\pi$. 
The $K^{*+}$ electric polarizability appears to be negative, just as we observed for the $K^{* 0}$.

%
%
%
%
%

\section{Final Words}

The response of hadrons to applied electromagnetic fields gives us a 
probe of the underlying QCD dynamics. Electromagnetic 
multipole polarizabilities encode the longest-range response to the external 
field. As such, these observables are highly constrained by chiral 
dynamics, and have been the subject of a continuing experimental effort. 
Lattice QCD can provide first principles determination of polarizabilities. 
We have briefly reviewed our on-going background field calculations,
focusing on the electric polarizabilities of hadrons from a hybrid lattice action. 
Measurements for both neutral and charged particles 
are possible, the latter by assessing the behavior of background field
two-point correlation functions. 
Our results are preliminary at this stage. There are clearly a number
of refinements that can be made. Handling the exceptional gauge 
configurations is most likely necessary to improve the statistics
on the lighter quark mass ensembles. 
Ultimately we must address the infinite volume and continuum limits, 
as well as turn on the electric charges of the sea quarks.

\begin{acknowledgments}
This work is supported in part by the 
U.S.~Deptartment~of Energy,
Grants 
No.~DE-FG02-97ER-41014
(W.D.),
No.~DE-FG02-93ER-40762
(B.C.T. and A.W.-L.),
and
No.~DE-FG02-07ER-41527
(A.W.-L.).
\end{acknowledgments}


\begin{thebibliography}{9}



\bibitem{Hu:2007eb}
  J.~Hu, F.-J.~Jiang and B.~C.~Tiburzi,
  Phys.\ Lett.\  B {\bf 653}, 350 (2007);
  Phys.\ Rev.\  D {\bf 77}, 014502 (2008).



\bibitem{HydeWright:2004gh}
  C.~E.~Hyde and K.~de Jager,
  Ann.\ Rev.\ Nucl.\ Part.\ Sci.\  {\bf 54}, 217 (2004).
  


\bibitem{Allton:2007hx}
  C.~Allton {\it et al.}  [RBC and UKQCD Collaborations],
  Phys.\ Rev.\  D {\bf 76}, 014504 (2007);
  arXiv:0804.0473 [hep-lat].


\bibitem{Tiburzi:2008ma}
  B.~C.~Tiburzi,
  arXiv:0808.3965 [hep-ph].



\bibitem{'tHooft:1979uj}
  G.~'t Hooft,
  Nucl.\ Phys.\  B {\bf 153}, 141 (1979);
  P.~van Baal,
  Commun.\ Math.\ Phys.\  {\bf 85}, 529 (1982).

\bibitem{Smit:1986fn}
  J.~Smit and J.~C.~Vink,
  Nucl.\ Phys.\  B {\bf 286}, 485 (1987);
  H.~R.~Rubinstein, S.~Solomon and T.~Wittlich,
  Nucl.\ Phys.\  B {\bf 457}, 577 (1995).




\bibitem{Detmold:2006vu}
  W.~Detmold, B.~C.~Tiburzi and A.~Walker-Loud,
  Phys.\ Rev.\  D {\bf 73}, 114505 (2006). 



\end{thebibliography}
\end{document}